\documentclass[aps,prd,reprint,superscriptaddress,]{revtex4-2}

\usepackage{color}
\usepackage{graphicx}
\usepackage{amsmath}

\begin{document}

\title{Beyond Half-Life Limits: Robust Operator-Level Interpretation of Multi-Isotope Neutrinoless Double-Beta Decay}

\newcommand{\tohoku}{\affiliation{Research Center for Neutrino
    Science, Tohoku University, Sendai, Miyagi 980-8578, Japan}}

\author{K.~Ishidoshiro}\tohoku



\begin{abstract}
Current and future neutrinoless double-beta decay ($0\nu\beta\beta$) searches are usually characterized by half-life limits or projected sensitivities. For operator-level interpretation, however, the relevant question is not only the strength of the limit, but also the stability of the inferred constraints on low-energy lepton-number-violating~(LNV) coefficients under changes of isotope and nuclear matrix-element~(NME) method.

Using \texttt{$\nu$DoBe}, we convert half-life limits or projected sensitivities into limits on low-energy LNV coefficients for several choices of isotope, operator class, and NME method. We then compare the resulting limits across isotopes and NME methods. We quantify the stability by the spread of coefficient limits in non-observation scenarios and, for single-operator post-discovery benchmarks, by the spread of isotope-to-isotope half-life ratios.

For the operator examples considered in this paper, the light-Majorana exchange mechanism gives the most stable interpretation. The selected dimension-six operators remain reasonably robust, while the short-range dimension-nine examples show a stronger dependence on isotope and NME-method choices. The same hierarchy appears in post-discovery benchmarks based on isotope-to-isotope half-life ratios: the reliability of such tests depends on whether the predicted half-life pattern remains stable across NME methods.

A multi-isotope $0\nu\beta\beta$ program therefore provides information beyond half-life reach by testing the stability of the operator interpretation. This spread-based analysis provides a practical way to identify which low-energy operator scenarios are suitable for isotope-ratio tests after a possible discovery, and which require more care before being connected to ultraviolet models of lepton-number violation.

\end{abstract}


\maketitle

\section{Introduction}
\label{sec:intro}

Neutrinoless double-beta decay ($0\nu\beta\beta$) is one of the most sensitive probes of lepton-number violation (LNV). If observed, it would establish the existence of a $\Delta L=2$ process and, through the black-box theorem of Schechter and Valle~\cite{Schechter:1981bd}, imply that neutrinos have a Majorana mass component~\cite{Agostini:2022zub,GomezCadenas:2023sua,Cirigliano:2018yza}. This makes $0\nu\beta\beta$ searches a central part of the experimental program in neutrino and astroparticle physics.

The interpretation of a possible signal, or even of a null result, is not unique. The standard light-Majorana neutrino exchange mechanism is often used as a reference case, but many other long-range and short-range LNV operators can contribute in a low-energy effective-field-theory~(EFT) description~\cite{Cirigliano:2017djv,Cirigliano:2018yza,Scholer:2023ykx,Graf:2022rco}. A half-life limit therefore does not constrain only one parameter. It constrains a space of possible LNV interactions, with the mapping depending on the assumed operator and on the nuclear input.

Future $0\nu\beta\beta$ searches will improve the half-life reach, but this is not the only relevant change. They will also provide results from different isotopes, experimental techniques, and nuclear matrix-element~(NME) calculations. This raises a practical question: once a half-life limit or sensitivity is translated into a bound on a low-energy LNV coefficient, how stable is that bound under changes of isotope and NME method?

This is not the same as asking which isotope gives the strongest limit. The strongest isotope can depend on the operator. Moreover, a strong best limit does not necessarily imply a stable interpretation if the isotope pattern changes substantially across NME methods. Conversely, an operator with a less aggressive absolute reach may still be easier to interpret if its multi-isotope pattern is more stable.

The use of several isotopes has long been discussed as a way to extract more information than a single half-life can provide. Different nuclei involve different phase-space factors, NMEs, and hadronic inputs, so isotope-to-isotope comparisons can in principle help distinguish among underlying mechanisms. Characteristic half-life patterns have been studied in this context~\cite{Deppisch:2006hb,Gehman:2007qg}, and recent work has also emphasized the role of multiple isotopes in resolving operator- and model-level degeneracies~\cite{Agostini:2022zub,Chen:2024fue}. Related EFT-based studies of nonstandard $0\nu\beta\beta$ mechanisms have used the low-energy operator framework to study mechanism discrimination and the associated nuclear and hadronic inputs~\cite{Graf:2022rco}.

The usefulness of such comparisons, however, depends on whether the predicted patterns are stable against variations in the nuclear input, including NME uncertainties~\cite{Engel:2016xgb}. This point is important because changing the isotope changes not only the experimental sensitivity, but also the phase-space factor, the relevant matrix elements, and the relative weight of different operator structures. Different NME methods can then shift the inferred coefficient limits in an operator-dependent way. For this reason, comparing half-life sensitivities alone is not enough to judge the robustness of the operator interpretation.

Here we ask a slightly earlier question: are the operator patterns used in such comparisons stable enough against isotope and NME-method variations to make the comparison reliable? Rather than attempting a global uncertainty budget or a full mechanism-discrimination analysis, we use the low-energy operator information to ask which operator patterns remain stable enough across a representative multi-isotope program to support reliable interpretation. The emphasis is therefore on the stability of the interpretation, rather than on distinguishability alone. We refer to this restricted notion of stability as operator-level interpretability: the stability of the inferred low-energy operator information under variations of isotope and NME method. This restricted definition concerns the robustness of the inference within a fixed operator hypothesis, and should not be confused with the distinguishability of different operator hypotheses. A large isotope dependence may still be diagnostically useful if it produces a distinctive and theoretically stable isotope pattern. In this terminology, the spread obtained by varying both isotope and NME method is not meant to be a pure NME uncertainty. It is a benchmark-set diagnostic of interpretive stability. The corresponding NME-only and isotope-only decompositions are given separately in
Appendix~\ref{app:spread_decomposition}.

We study this question using current and future benchmark isotope sets in a common \texttt{$\nu$DoBe}-based framework. For each isotope, operator, and NME method, we translate the half-life sensitivity into a limit on the corresponding low-energy coefficient. We then compare these limits across isotopes and nuclear methods. As simple quantitative diagnostics of this stability, we use the spread of coefficient limits in non-observation scenarios and, for post-discovery benchmarks with single-operator dominance, the spread of isotope-to-isotope half-life ratios across NME methods.

\section{Framework}
\label{sec:framework}

In this section, we specify the representative isotope sets, the coefficient-limit construction, and the diagnostics used in the non-observation and post-discovery analyses.

\subsection{Representative isotope sets}
\label{subsec:benchmarks}

We consider one current and one future representative isotope set. These inputs are not intended to provide a complete survey of existing and planned searches. They are chosen to capture the characteristic information content of present and prospective multi-isotope programs within a compact benchmark setup.

For the current representative set, we adopt three isotopes:
\begin{align}
{}^{136}\mathrm{Xe}: \quad & T_{1/2}^{0\nu} > 3.8 \times 10^{26}\ \mathrm{yr}, \\
{}^{76}\mathrm{Ge}: \quad & T_{1/2}^{0\nu} > 1.9 \times 10^{26}\ \mathrm{yr}, \\
{}^{130}\mathrm{Te}: \quad & T_{1/2}^{0\nu} > 3.2 \times 10^{25}\ \mathrm{yr}.
\end{align}
For $^{136}$Xe, we use the complete KamLAND-Zen result~\cite{Abe:2024eml}. For $^{76}$Ge, we use the combined result from GERDA, the Majorana Demonstrator, and LEGEND-200~\cite{Acharya:2025legend}. For $^{130}$Te, we use the CUORE limit~\cite{Adams:2020qqz}.

For the future representative set, we adopt
\begin{align}
{}^{136}\mathrm{Xe}: \quad & T_{1/2}^{0\nu} = 2.0 \times 10^{27}\ \mathrm{yr}, \\
{}^{76}\mathrm{Ge}: \quad & T_{1/2}^{0\nu} = 1.0 \times 10^{28}\ \mathrm{yr}, \\
{}^{100}\mathrm{Mo}: \quad & T_{1/2}^{0\nu} = 1.6 \times 10^{27}\ \mathrm{yr}.
\end{align}
For $^{136}$Xe, we use a target sensitivity of $2\times10^{27}$ yr for KamLAND2-Zen. For $^{76}$Ge, we adopt the LEGEND-1000 design sensitivity beyond $10^{28}$ yr and use $1.0\times10^{28}$ yr as a representative value~\cite{Abgrall:2021rek}. For $^{100}$Mo, we use the baseline CUPID exclusion sensitivity of $1.6\times10^{27}$ yr~\cite{Alfonso:2025cupid}.

This future set is chosen to represent the major next-generation xenon, germanium, and molybdenum directions with complementary experimental and nuclear systematics. The replacement of $^{130}$Te by $^{100}$Mo is a benchmark choice reflecting the adopted CUPID sensitivity, not a statement that Te-based searches are intrinsically less important.

The current and future sets should be understood as benchmark information sets. The comparison below is meant to test operator-dependent complementarity in a prospective multi-isotope setting, rather than to give a literal chronological extrapolation.

\subsection{Operator basis, coefficient limits, and diagnostics}
\label{subsec:limits}

We use \(\nu\)DoBe~\cite{Scholer:2023ykx} to evaluate $0\nu\beta\beta$ rate factors for each isotope, operator, and NME method. \(\nu\)DoBe is a public code that implements the low-energy EFT description of $0\nu\beta\beta$ and provides isotope-, operator-, and nuclear-input-dependent rate factors in a common convention. This gives a common low-energy basis for all operator classes considered here and allows isotope complementarity and method dependence to be compared on the same footing. 
We consider four NME methods: CDFT, IBM2, QRPA, and SM. These labels denote the \texttt{$\nu$DoBe} NME sets based on covariant density functional theory~\cite{Ding:2024cft}, the interacting boson model~\cite{Deppisch:2020ztt}, the quasiparticle random-phase approximation~\cite{Hyvarinen:2015bda}, and the nuclear shell model~\cite{Menendez:2018jp}, respectively.

The full low-energy \texttt{$\nu$DoBe} basis contains many operator structures. For the main analysis, we focus on a compact set that covers the main physical categories relevant here: the standard light-Majorana exchange mechanism, selected long-range dimension-six nonstandard operators, and short-range dimension-nine examples. We use
\begin{equation}
m_{\beta\beta},\quad V_L^{(6)},\quad T^{(6)},\quad
3L^{(9)},\quad 6^{(9)}
\end{equation}
as the illustrative operator set. These representative coefficients and their roles in the present analysis are summarized in Table~\ref{tab:operator_definitions}. The purpose of this choice is to sample qualitatively different interpretability regimes, not to rank operators or to cover the full low-energy basis. The coefficient $m_{\beta\beta}$ gives the standard light-Majorana reference case. The operators $V_L^{(6)}$ and $T^{(6)}$ represent selected long-range dimension-six vector and tensor structures, while $3L^{(9)}$ and $6^{(9)}$ serve as short-range dimension-nine examples in the \texttt{$\nu$DoBe} basis. The operator $6^{(9)}$ is included to represent a dimension-nine spread pattern that is distinct from the $3L^{(9)}$ pattern in the full survey. 

\begin{table*}[t]
\centering
\caption{Representative low-energy coefficients used in the main analysis. The labels follow the \texttt{$\nu$DoBe} convention of Ref.~\cite{Scholer:2023ykx}. The explicit low-energy operator definitions are those of Ref.~\cite{Scholer:2023ykx}. Coefficients of different operator dimensions have different mass dimensions and are not compared directly as dimensionless quantities.}
\label{tab:operator_definitions}
\begin{ruledtabular}
\begin{tabular}{lll}
Coefficient & Class & Role in this work \\
$m_{\beta\beta}$ & light-Majorana exchange & standard reference mechanism \\
$V_L^{(6)}$ & long-range dimension-six vector coefficient & selected dimension-six example \\
$T^{(6)}$ & long-range dimension-six tensor coefficient & selected dimension-six example \\
$3L^{(9)}$ & short-range dimension-nine coefficient & moderate-spread short-range example \\
$6^{(9)}$ & short-range dimension-nine coefficient & large-spread short-range example \\
\end{tabular}
\end{ruledtabular}
\end{table*}

Throughout this work, coefficient limits are obtained by switching on one low-energy operator at a time. Interference among different operator classes is therefore not included in the limits quoted in this work. The operator labels and coefficient normalizations follow the low-energy \texttt{$\nu$DoBe} basis, and all coefficients should be interpreted within this convention. In particular, coefficients associated with different operator dimensions carry different mass dimensions and should not be compared directly as dimensionless quantities.

For the standard light-Majorana mechanism, the coefficient $m_{\beta\beta}$ has mass dimension one and is quoted in GeV. Numerically, the resulting limits correspond to the usual meV-scale effective Majorana mass range. More precisely, the $m_{\beta\beta}$ parameter used here should be understood in the \texttt{$\nu$DoBe} normalization of the light-Majorana exchange contribution, and corresponds to the standard effective Majorana mass parameter up to convention-dependent normalization factors.

We use natural units with $\hbar=c=1$. For a given isotope, operator, and NME method, collectively denoted by $x$, we write the inverse half-life schematically as
\begin{equation}
\left[T_{1/2}^{0\nu}\right]^{-1} = K_x |C|^2,
\label{eq:rate_definition}
\end{equation}
where $C$ is the relevant low-energy effective coefficient and $K_x$ is the corresponding rate factor. In the notation used here, $K_x$ includes the phase-space, nuclear-matrix-element, and hadronic matching inputs adopted in \texttt{$\nu$DoBe}. The half-life sensitivity can therefore be translated into a coefficient limit according to
\begin{equation}
|C|_{\max} = \sqrt{\frac{1}{K_x T_{1/2}}}.
\end{equation}
This coefficient limit is the basic quantity used in the non-observation analysis. Here the subscript ``max'' denotes the maximum value of the coefficient allowed by a given half-life limit or sensitivity, not a maximum over isotope or NME-method choices.

For each operator class, we define the best and worst limits across all isotope $\times$ method combinations. The best limit is the strongest coefficient limit, while the worst limit is the weakest coefficient limit. To quantify the gain from the future set, we define
\begin{equation}
\mathrm{improvement}
\equiv
\frac{|C|^{\mathrm{current}}_{\max}}{|C|^{\mathrm{future}}_{\max}},
\end{equation}
using the best limit in each scenario. This quantity measures the improvement in the strongest coefficient reach. It does not, by itself, measure how stable the corresponding operator interpretation is.

To compare the stability of different operator classes, we define the spread of extracted coefficient limits across isotope and NME-method combinations,
\begin{equation}
S_{\rm bench}(O) \equiv
\frac{|C|_{\max}^{\mathrm{worst}}}{|C|_{\max}^{\mathrm{best}}}.
\end{equation}
Here the best and worst limits are taken over all isotope and NME-method combinations in the adopted benchmark set for a fixed operator \(O\). We therefore refer to this quantity as the benchmark-set spread. Smaller spread indicates that the inferred low-energy information is more stable within the adopted set of isotopes and NME methods. This quantity is not intended as a statistical uncertainty estimator, because we do not assign probabilities to the isotope choices or to the different NME methods. Instead, it is a worst-case robustness diagnostic: it measures how much the extracted coefficient limit can change within the adopted benchmark set when the isotope and NME method are varied. Because both isotope and NME method are varied, the benchmark-set spread should not be interpreted as a pure nuclear-theory uncertainty. It also contains the isotope dependence of the phase-space factors, the matrix-element normalization, and the adopted experimental sensitivity pattern. We use it to classify operator scenarios according to the stability of their interpretation, rather than to infer a probability distribution for the coefficient.

To make this separation explicit, Appendix~\ref{app:spread_decomposition} also gives two decompositions of this quantity. The NME-only spread at fixed isotope is defined by varying only the NME method,
\begin{equation}
S_{\rm NME}(i,O)=
\frac{\max_{\alpha}|C|_{\max}(i,O,\alpha)}
{\min_{\alpha}|C|_{\max}(i,O,\alpha)},
\end{equation}
where \(i\) labels the isotope and \(\alpha\) the NME method. Conversely, the isotope-only spread at fixed NME method is defined as
\begin{equation}
S_{\rm iso}(\alpha,O)=
\frac{\max_i |C|_{\max}(i,O,\alpha)}
{\min_i |C|_{\max}(i,O,\alpha)}.
\end{equation}
These decompositions are used only to diagnose the origin of the benchmark-set spread; the main figures retain \(S_{\rm bench}\) because the goal is to assess the stability of the interpretation within a representative multi-isotope program.

We check the compact operator set against the full operator survey in Appendix~\ref{app:full_operator_survey}. As shown there, the selected operators cover the main patterns found in the benchmark analysis: a comparatively stable standard case, moderately robust dimension-six examples, and dimension-nine examples with stronger dependence on the nuclear treatment.

\subsection{Post-discovery benchmark setup}
\label{subsec:discoveryframework}

We also consider post-discovery benchmark scenarios under single-operator dominance. In these scenarios, one reference isotope is assumed to fix the coefficient for a given operator hypothesis. The relevant observables are then the isotope-to-isotope half-life ratios predicted for the remaining nuclei. As in the non-observation analysis, we characterize the stability of these predictions by a \emph{ratio spread}. This setup is used to test whether the post-discovery isotope pattern remains stable across NME methods. 

\section{Non-observation constraints from representative multi-isotope sets}
\label{sec:nonobs}

We translate the adopted half-life sensitivities into limits on low-energy LNV coefficients for each isotope, operator class, and NME method. We focus on the representative operator set; the full operator-by-operator survey is shown in Appendix~\ref{app:full_operator_survey}.

\subsection{Current and future representative-set constraints}

We first evaluate the current set consisting of $^{136}$Xe, $^{76}$Ge, and $^{130}$Te, and then compare it with the future set consisting of $^{136}$Xe, $^{76}$Ge, and $^{100}$Mo. For each coefficient, we summarize the best and worst limits obtained across all isotope$\times$NME-method combinations. 

The results for the compact operator set are shown in Table~\ref{tab:representative_summary} and Fig.~\ref{fig:representative_summary}. The table lists the best current limit, the best and worst future limits, the future benchmark-set spread, the current-to-future improvement factor, and the isotope giving the best projected limit. The figure shows the corresponding best coefficient limits and benchmark-set spreads for the current and future representative sets.



\subsection{Improvement factors and isotope complementarity}
\label{subsec:improvement}

We first discuss the improvement factor, which quantifies the gain in the strongest coefficient reach from the current to the future set. The relevant values are listed in Table~\ref{tab:representative_summary}, with coefficient normalizations following the \texttt{$\nu$DoBe} conventions.

Although the improvement factors for these operators are all of order three, Table~\ref{tab:representative_summary} shows that the isotope giving the best projected limit differs by operator class.  We obtain
\begin{equation}
\begin{gathered}
m_{\beta\beta}: 3.21,\qquad
T^{(6)}: 3.36,\qquad
V_L^{(6)}: 3.40,\\
3L^{(9)}: 3.36,\qquad
6^{(9)}: 3.09.
\end{gathered}
\end{equation}
The overall size follows the expected square-root scaling with half-life sensitivity, while the isotope pattern remains operator dependent. In particular, the isotope giving the strongest projected limit depends on the operator class. 

This operator dependence is physically important. The future set is not simply a uniform rescaling of the current sensitivity pattern. Rather, it changes the structure of the operator-space information by altering which isotope dominates the best limit. The stability of the inferred coefficient constraint is discussed separately below in terms of the benchmark-set spread.

\subsection{Spread and robustness of interpretation}

\begin{figure*}[t]
  \centering
  \includegraphics[width=0.88\textwidth]{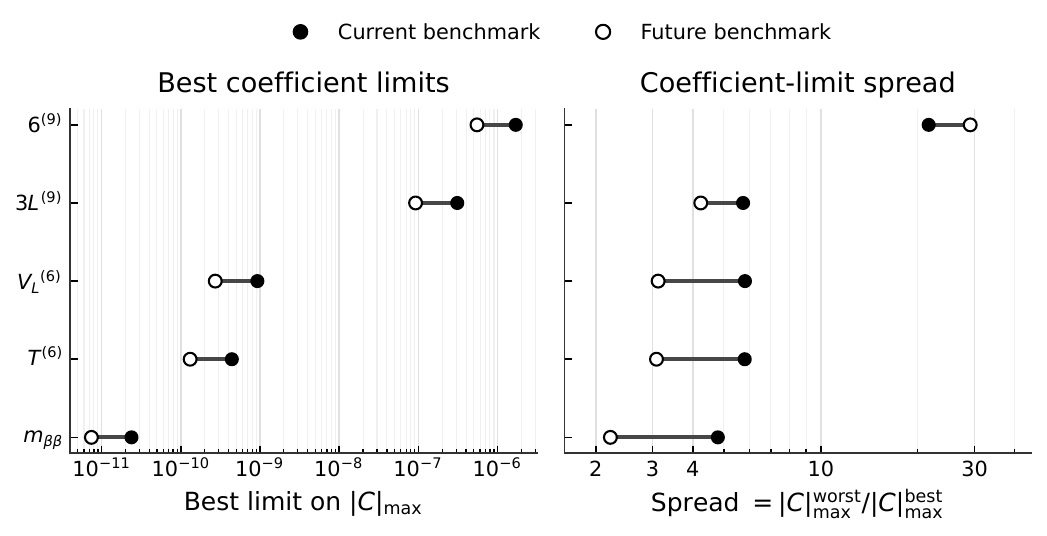}
  \caption{
  Summary of the compact operator set. The left panel shows the best coefficient limits in the current and future sets. The right panel shows the benchmark-set spread of coefficient limits across isotope and NME-method combinations. Smaller spread indicates greater interpretive stability within the adopted benchmark set, but the quantity should not be read as a pure NME uncertainty. The full operator-by-operator survey is shown in Appendix~\ref{app:full_operator_survey}.
  }
  \label{fig:representative_summary}
\end{figure*}

We now turn to the benchmark-set coefficient-limit spread for the current and future isotope sets. The resulting values for the compact operator set are shown in the right panel of Fig.~\ref{fig:representative_summary}; the full operator-by-operator spread distribution is shown in Fig.~\ref{fig:app_spread} of Appendix~\ref{app:full_operator_survey}. Smaller values indicate that the inferred operator-level information remains comparatively stable under changes of isotope and NME method, whereas larger values signal stronger benchmark-set dependence and reduced interpretive robustness.

We use this number only for comparison: it is not a complete uncertainty estimate, but a practical worst-case classifier for comparing the robustness of different operator classes within the adopted set of isotopes and NME methods. In particular, because it varies both isotope and NME method, it should not be interpreted as a pure nuclear-theory uncertainty. Its purpose is to identify operator scenarios whose interpretation is stable enough to support multi-isotope comparisons, not to assign statistical weights to different nuclear calculations. The separation into NME-only and isotope-only components is given in Appendix~\ref{app:spread_decomposition}.

For the compact operator set, the future-set spreads are
\begin{equation}
\begin{gathered}
m_{\beta\beta}: 2.22,\qquad
T^{(6)}: 3.09,\qquad
V_L^{(6)}: 3.13,\\
3L^{(9)}: 4.24,\qquad
6^{(9)}: 29.09.
\end{gathered}
\end{equation}
The reduction of the spread in the future set should not be interpreted as a generic consequence of improving all half-life sensitivities. A uniform rescaling of all isotope sensitivities would leave the spread unchanged. The smaller future spreads for the light-Majorana and selected dimension-six cases arise because the adopted future benchmark has a more balanced isotope reach: the relatively weak end of the current set is replaced by the next-generation \(^{100}\)Mo benchmark, while \(^{76}\)Ge also gains substantially. As a result, the separation between the best and worst isotope--method combinations is reduced for several operator classes. The much larger value for $6^{(9)}$, however, shows that this balancing of experimental reach does not remove operator-dependent nuclear-method sensitivity in all short-range cases. The remaining spread, especially for the dimension-nine examples, therefore reflects residual operator-dependent nuclear-method variation rather than only experimental reach.

The standard light-Majorana case is therefore the most stable among this set. The dimension-six classes $T^{(6)}$ and $V_L^{(6)}$ occupy an intermediate but still comparatively robust position. The two representative dimension-nine operators illustrate that the short-range sector is not homogeneous in the present benchmark setup: $3L^{(9)}$ gives a moderate-spread example, whereas $6^{(9)}$ gives a large-spread example.

These results confirm that the compact operator set spans the main interpretability patterns found in the non-observation analysis.

\begin{table*}[t]
\centering
\caption{
Summary of current-to-future non-observation constraints for the compact operator set. Best and worst limits are taken over all isotope$\times$method combinations. The quoted spread is the benchmark-set spread defined in Sec.~\ref{subsec:limits}. The improvement factor is defined in Sec.~\ref{subsec:limits}. All coefficients are quoted in the \texttt{$\nu$DoBe} normalization. For $m_{\beta\beta}$, the best current value corresponds to the $^{136}$Xe--CDFT entry in the \texttt{$\nu$DoBe} input and represents the optimistic edge of the internally consistent benchmark.
}
\label{tab:representative_summary}
\setlength{\tabcolsep}{4pt}
\renewcommand{\arraystretch}{1.08}
\begin{tabular}{lcccccc}
\hline
Coeff. & Best$_{\rm cur}$ & Best$_{\rm fut}$ & Worst$_{\rm fut}$ & Spread$_{\rm fut}$ & Impr. & Best iso.$_{\rm fut}$ \\
\hline
$m_{\beta\beta}$ & $2.39\times 10^{-11}$ & $7.44\times 10^{-12}$ & $1.65\times 10^{-11}$ & 2.22 & 3.21 & $^{100}$Mo \\
$T^{(6)}$ & $4.41\times 10^{-10}$ & $1.31\times 10^{-10}$ & $4.05\times 10^{-10}$ & 3.09 & 3.36 & $^{76}$Ge \\
$V_L^{(6)}$ & $9.26\times 10^{-10}$ & $2.72\times 10^{-10}$ & $8.51\times 10^{-10}$ & 3.13 & 3.40 & $^{76}$Ge \\
$3L^{(9)}$ & $3.09\times 10^{-7}$ & $9.21\times 10^{-8}$ & $3.9\times 10^{-7}$ & 4.24 & 3.36 & $^{100}$Mo \\
$6^{(9)}$ & $1.69\times 10^{-6}$ & $5.49\times 10^{-7}$ & $1.6\times 10^{-5}$ & 29.09 & 3.09 & $^{100}$Mo \\
\hline
\end{tabular}
\end{table*}

\begin{table*}[tbp]
\centering
\caption{Representative isotope-to-isotope half-life ratios in the post-discovery reference scenarios. For all selected operator classes, $^{136}$Xe is taken as the reference isotope. The table lists the median ratio across NME methods together with the ratio spread, defined as the maximum-to-minimum ratio across methods. Large ratio spread indicates that the predicted isotope pattern is more sensitive to the nuclear method and therefore has reduced diagnostic stability.}
\label{tab:discovery_ratios}
\begin{tabular}{lcccc}
\hline
Coefficient & Reference isotope & Target isotope & Median $T_{\rm target}/T_{\rm ref}$ & Ratio spread \\
\hline
$m_{\beta\beta}$ & $^{136}$Xe & $^{100}$Mo & 0.395 & 1.22 \\
$m_{\beta\beta}$ & $^{136}$Xe & $^{76}$Ge & 2.899 & 2.93 \\
$T^{(6)}$ & $^{136}$Xe & $^{100}$Mo & 0.467 & 1.22 \\
$T^{(6)}$ & $^{136}$Xe & $^{76}$Ge & 3.197 & 2.70 \\
$V_L^{(6)}$ & $^{136}$Xe & $^{100}$Mo & 0.478 & 1.22 \\
$V_L^{(6)}$ & $^{136}$Xe & $^{76}$Ge & 3.122 & 2.70 \\
$3L^{(9)}$ & $^{136}$Xe & $^{100}$Mo & 0.385 & 4.95 \\
$3L^{(9)}$ & $^{136}$Xe & $^{76}$Ge & 3.122 & 2.47 \\
$6^{(9)}$ & $^{136}$Xe & $^{100}$Mo & 0.078 & 202.58 \\ 
$6^{(9)}$ & $^{136}$Xe & $^{76}$Ge & 3.976 & 716.11 \\
\hline
\end{tabular}
\end{table*}

\section{POST-DISCOVERY HALF-LIFE RATIO TESTS}
\label{sec:discovery}

We now turn to the discovery regime and ask how stably an observed half-life in one isotope predicts the corresponding expectations in other isotopes under the same operator hypothesis. Throughout this section, we work within a controlled benchmark framework based on single-operator dominance.

\subsection{Benchmark setup and half-life ratios}

For definiteness, we use $^{136}$Xe to organize the comparison, without assuming that Xe is universally optimal.  The post-discovery question is the following: if a signal in $^{136}$Xe fixes the coefficient of a given single-operator hypothesis, what half-life would the same hypothesis predict in another isotope?  The ratios discussed in this section are therefore not ratios of the adopted experimental sensitivities, but predicted half-life ratios for a fixed operator and a common coefficient.

Using Eq.~\eqref{eq:rate_definition}, the ratio for a fixed operator $O$ and NME method $\alpha$ is
\begin{equation}
\frac{T_{\mathrm{target}}}{T_{\mathrm{ref}}}
=
\frac{K_{\mathrm{ref},\alpha}^{(O)}}
     {K_{\mathrm{target},\alpha}^{(O)}} .
\end{equation}
Thus the ratios are operator- and method-dependent, because the rate factors $K_{i,\alpha}^{(O)}$ depend on the isotope $i$, operator $O$, and NME method $\alpha$. 

To quantify the method dependence of each predicted ratio, we define
\begin{equation}
\mathrm{ratio\ spread}
\equiv
\frac{(T_{\mathrm{target}}/T_{\mathrm{ref}})_{\max}}
     {(T_{\mathrm{target}}/T_{\mathrm{ref}})_{\min}},
\end{equation}
where the maximum and minimum are taken over the adopted NME methods. Smaller ratio spread indicates that the predicted isotope pattern is more stable against the choice of nuclear method.

The results are summarized in Table~\ref{tab:discovery_ratios}, where the quoted half-life ratios are medians over the adopted NME methods. The standard light-Majorana case $m_{\beta\beta}$ remains comparatively stable: with $^{136}$Xe as the reference isotope, we find $T_{^{100}\mathrm{Mo}}/T_{^{136}\mathrm{Xe}} = 0.395$ with ratio spread $1.22$, and $T_{^{76}\mathrm{Ge}}/T_{^{136}\mathrm{Xe}} = 2.90$ with ratio spread $2.93$.

The dimension-six classes also show relatively controlled behavior. For $T^{(6)}$, the median predicted ratios are $T_{^{100}\mathrm{Mo}}/T_{^{136}\mathrm{Xe}} = 0.467$ and $T_{^{76}\mathrm{Ge}}/T_{^{136}\mathrm{Xe}} = 3.20$, with ratio spreads $1.22$ and $2.70$, respectively. For $V_L^{(6)}$, the corresponding median ratios are $0.478$ and $3.12$, with ratio spreads $1.22$ and $2.70$. These classes therefore retain meaningful post-discovery diagnostic power, though they are somewhat less stable than the standard case.

The two short-range dimension-nine examples show different levels of post-discovery stability. For $3L^{(9)}$, the median predicted ratios are $T_{^{100}\mathrm{Mo}}/T_{^{136}\mathrm{Xe}} = 0.385$ and $T_{^{76}\mathrm{Ge}}/T_{^{136} \mathrm{Xe}} = 3.12$, with ratio spreads $4.95$ and $2.47$, respectively. For $6^{(9)}$, the corresponding median ratios  are $0.078$ and $3.98$, but the ratio spreads increase to $202.58$ and $716.11$. The latter case therefore illustrates a large-spread dimension-nine pattern in which isotope-ratio predictions become highly sensitive to the adopted nuclear method.

\subsection{Diagnostic stability of isotope-ratio tests}

Table~\ref{tab:discovery_ratios} shows a clear hierarchy, but also shows that the dimension-nine sector is not represented by a single level of instability in the present benchmark setup. The standard light-Majorana case has the smallest spreads, the selected dimension-six classes remain moderately stable, $3L^{(9)}$ gives a moderate-spread short-range example, and $6^{(9)}$ gives a much larger-spread example.

The hierarchy also depends on the isotope pair. For $m_{\beta\beta}$, the dimension-six examples, and $3L^{(9)}$, the $^{76}$Ge-to-$^{136}$Xe and $^{100}$Mo-to-$^{136}$Xe ratio spreads are of order a few or smaller. By contrast, $6^{(9)}$ gives very large ratio spreads for both isotope pairs, especially for the $^{76}$Ge-to-$^{136}$Xe comparison. This shows that post-discovery diagnostic power is both operator-dependent and isotope-pair dependent.

Comparison across isotopes contains more information than a single-isotope observation, but its usefulness is operator dependent. For some classes it provides a robust bridge from discovery to interpretation; for others it remains too method dependent to support strong conclusions. The ratio spread also gives a simple way to estimate when two operator hypotheses can be robustly separated by isotope-ratio information. Suppose two operator hypotheses $A$ and $B$ predict median isotope-to-isotope half-life ratios $R_A$ and $R_B$, with corresponding ratio spreads $S_A$ and $S_B$. If their separation is not larger than the method-induced variation,
\begin{equation}
\frac{\max(R_A,R_B)}{\min(R_A,R_B)}
\lesssim
\max(S_A,S_B),
\end{equation}
then the apparent difference between the two hypotheses is not robust against the choice of nuclear method. In this sense, the ratio spread quantifies the minimum separation required for a stable isotope-ratio diagnostic, before experimental uncertainties and mixed-operator effects are included.

This criterion gives a simple semi-quantitative measure of the risk of misinterpretation in a post-discovery setting. A ratio spread close to unity implies that the predicted isotope pattern is stable and can provide genuine diagnostic power, whereas a spread of order several means that the prediction itself varies by a factor of several. In the present benchmarks, the $^{100}$Mo-to-$^{136}$Xe ratio spread is $4.95$ for $3L^{(9)}$, so the predicted ratio changes by nearly a factor of five across the adopted NME methods. For $6^{(9)}$, the variation is much larger, reaching $202.58$ for the $^{100}$Mo-to-$^{136}$Xe ratio and $716.11$ for the $^{76}$Ge-to-$^{136}$Xe ratio. In such cases, model selection or coefficient extraction based on isotope ratios alone would be vulnerable to method-induced bias, even before experimental uncertainties or mixed-operator effects are included.

\section{Discussion}
\label{sec:discussion}
\begin{figure}[t]
  \centering
  \includegraphics[width=0.9\linewidth]{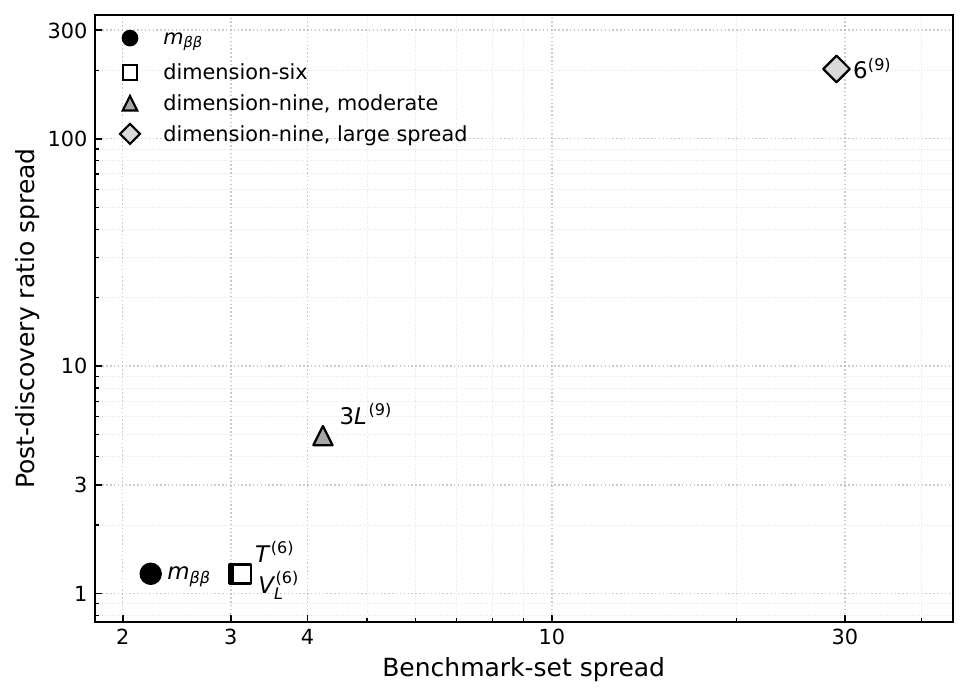}
\caption{
Interpretability plane for the selected operator classes. The horizontal axis shows the benchmark-set spread in the future representative set, and the vertical axis shows the post-discovery ratio spread for the $^{100}$Mo-to-$^{136}$Xe comparison. Smaller values on both axes indicate more stable interpretation. The light-Majorana mechanism lies in the most stable region, the selected dimension-six operators remain comparatively robust, while the two dimension-nine examples illustrate distinct spread patterns: $3L^{(9)}$ gives a moderate-spread case and $6^{(9)}$ gives a large-spread case in the adopted benchmark setup.
}
  \label{fig:interpretability_plane}
\end{figure}

The two diagnostics discussed above can be combined into an interpretability plane, defined here as the plane spanned by the benchmark-set spread in the non-observation analysis and the post-discovery ratio spread, as shown in Fig.~\ref{fig:interpretability_plane}. This representation makes clear that, within the present benchmark setup, $m_{\beta\beta}$ lies in the most stable region, the selected dimension-six operators remain reasonably robust, and the two dimension-nine examples illustrate different spread patterns. It also illustrates the main point of this work: a stronger coefficient reach does not necessarily imply a more stable operator interpretation.

The decompositions in Appendix \ref{app:spread_decomposition} further show that the benchmark-set spread is not driven by a single source: it combines NME-method variation at fixed isotope with isotope-to-isotope variation at fixed NME method. In this respect, the two selected dimension-nine examples serve complementary roles: \(3_L^{(9)}\) gives a moderate-spread pattern, whereas \(6^{(9)}\) exposes a large-spread pattern driven mainly by nuclear-method variation in the \(^{136}\mathrm{Xe}\) entries. 

It is important to distinguish the robustness quantified by the present spreads from operator distinguishability. The isotope-only contribution measures how strongly the extracted coefficient reach depends on the adopted isotope and sensitivity pattern; a large value does not by itself imply that the operator hypothesis is difficult to identify. Indeed, a strongly isotope-dependent pattern may improve discrimination if it is distinctive and remains stable across NME methods. Conversely, a small isotope dependence does not guarantee distinguishability if several operators predict similar isotope ratios.

There are several limitations to this comparison. The spread and ratio spread are defined only with respect to the isotope sets and NME methods included in this study. The resulting classification is therefore comparative, not absolute. The post-discovery benchmarks also assume single-operator dominance, and the current and future isotope sets are representative choices rather than a complete experimental forecast. Thus the spreads quoted here should not be read as full theory or nuclear-structure uncertainties. They are meant instead to provide a controlled way to identify which operator classes already have stable multi-isotope patterns before more complicated effects are included.

Another limitation is that the hadronic low-energy constants entering the chiral-EFT matching are not varied in this work. These constants encode nonperturbative QCD information in the mapping from quark-level LNV operators to pion- and nucleon-level interactions. Some of them are constrained by experiment or lattice-QCD calculations, while others are known only at the order-of-magnitude level. This is especially relevant for short-range dimension-nine operators, for which several hadronic couplings can enter the rate factors. The spreads quoted here should therefore be interpreted as diagnostics at fixed \texttt{$\nu$DoBe} hadronic input, not as a complete uncertainty budget including LEC variations. 

A combined variation of the NME and hadronic LEC inputs would be needed to determine which source dominates the theoretical uncertainty for each operator class. Such an analysis would also help identify whether future progress should primarily target nuclear-structure calculations or hadronic matching inputs, but is beyond the scope of the present work.

This single-operator baseline is still useful for thinking about more realistic cases. In ultraviolet (UV) completions, several low-energy operators may be generated at the same time, with model-dependent relative weights. Interference, cancellations, or degeneracies among these operators can then change both the total rate and the pattern of half-lives across isotopes. We refer to the set of induced low-energy operators and their relative weights as the operator footprint of the UV scenario. It is therefore more natural to regard a UV model as occupying a direction or region in low-energy operator space, rather than as a single point in the interpretability plane.

Operators that already show a large spread in the single-operator analysis are likely to be especially difficult to interpret once mixed-operator effects are allowed. Conversely, stable single-operator patterns provide a more favorable starting point for future mixed-operator and UV-matching analyses. A full reconstruction of the ultraviolet theory would require interference effects, renormalization-group running, and explicit model matching, and is left for future work.

\section{Conclusions}
\label{sec:conclusions}
We have developed a simple diagnostic framework for assessing the operator-level interpretability of multi-isotope $0\nu\beta\beta$ searches. Using representative current and future isotope sets in a common \texttt{$\nu$DoBe}-based setup, we converted half-life sensitivities into limits on low-energy operator coefficients and compared the results across isotopes and NME methods.

For the representative operators studied here, the light-Majorana mechanism gives the most stable interpretation. The selected dimension-six operators remain reasonably robust, while the short-range dimension-nine examples show more operator-dependent behavior, including cases with stronger sensitivity to nuclear-method choices. The representative set also illustrates that the dimension-nine sector is not characterized by a single level of stability in the adopted benchmark setup: $3L^{(9)}$ gives a moderate-spread short-range example, whereas $6^{(9)}$ gives a large-spread example. The same qualitative pattern also appears in the post-discovery benchmark study.

A multi-isotope program therefore carries information beyond the improvement of half-life sensitivity: it also tests the stability of the inferred operator information. In this sense, sensitivity reach and interpretive robustness should be treated as related but separate aspects of future $0\nu\beta\beta$ searches.

The present study should be regarded as a single-operator baseline. In a post-discovery setting, large ratio spreads would make isotope-pattern comparisons more vulnerable to NME-method dependence, and mixed-operator scenarios would introduce additional degeneracies. Applying the same diagnostics to mixed-operator cases and to explicit UV-matching scenarios would be the next step. 


\section*{Data Availability}
No new experimental data were generated in this work. The input half-life limits and projected sensitivities are taken from the published sources cited in Sec.~\ref{subsec:benchmarks}. The rate factors were evaluated with the publicly available \texttt{$\nu$DoBe} package, and the numerical values needed to reproduce the main results are given in the tables and figures.

\begin{acknowledgments}
This work was supported by JSPS KAKENHI Grant Numbers JP24H00209 and JP24H02237. 
\end{acknowledgments}

\clearpage
\appendix

\section{Decomposition of the benchmark-set spread}
\label{app:spread_decomposition}
The benchmark-set spread used in the main text varies both the isotope and the NME method. It is therefore useful to display separately the part induced by changing the NME method at fixed isotope and the part induced by changing the isotope at fixed NME method. This appendix gives this decomposition for the future representative set and for the compact operator set used in the main text.

Table~\ref{tab:app_nme_only_spread} shows the NME-only spread at fixed isotope. This quantity is most directly related to nuclear-method variation. The table shows that the size and isotope dependence of the NME-only spread are operator dependent. In particular, the large benchmark-set spread of $6^{(9)}$ is associated mainly with a large NME-method variation in the $^{136}$Xe entries, whereas the $^{100}$Mo and $^{76}$Ge entries are much more stable for this operator.

\begin{table*}[tbp]
\centering
\caption{NME-only spread at fixed isotope for the future representative set. The spread is defined as $S_{\rm NME}(i,O)=\max_\alpha |C|_{\max}(i,O,\alpha)/\min_\alpha |C|_{\max}(i,O,\alpha)$, where only the NME method $\alpha$ is varied.}
\label{tab:app_nme_only_spread}
\begin{tabular}{lccc}
\hline
Coefficient & $^{136}$Xe & $^{76}$Ge & $^{100}$Mo \\
\hline
$m_{\beta\beta}$ &1.59 & 2.07 & 1.56 \\
$T^{(6)}$  & 2.11  & 2.15 & 2.32 \\
$V_L^{(6)}$ & 2.11  & 2.15 & 2.32 \\
$3L^{(9)}$  & 1.91  & 2.56 & 4.24 \\
$6^{(9)}$   & 21.61 & 2.60 & 1.54 \\
\hline
\end{tabular}
\end{table*}

Table~\ref{tab:app_isotope_only_spread} shows the complementary isotope-only spread at fixed NME method. This quantity is not a nuclear-theory uncertainty by itself, because it also contains the isotope dependence of phase-space factors, matrix elements, and the adopted experimental sensitivities. It should therefore not be interpreted as a direct measure of operator distinguishability. Instead, it shows how much of the benchmark-set spread can arise from isotope and sensitivity choices once a particular NME method is held fixed.

\begin{table*}[tbp]
\centering
\caption{Isotope-only spread at fixed NME method for the future representative set. The spread is defined as $S_{\rm iso}(\alpha,O)=\max_i |C|_{\max}(i,O,\alpha)/\min_i |C|_{\max}(i,O,\alpha)$, where only the isotope $i$ is varied.}
\label{tab:app_isotope_only_spread}
\begin{tabular}{lcccc}
\hline
Coefficient & CDFT & IBM2 & QRPA & SM \\
\hline
$m_{\beta\beta}$ & 1.40 & 1.49 & 1.71 & 1.00 \\
$T^{(6)}$ & 1.31 & 1.58 & 1.47 & 1.04 \\
$V_L^{(6)}$ & 1.29 & 1.60 & 1.48 & 1.02 \\
$3L^{(9)}$ & 1.46 & 2.48 & 1.49 & 1.04 \\
$6^{(9)}$ & 1.32 & 3.20 & 23.60 & 1.13 \\
\hline
\end{tabular}
\end{table*}

As a by-product of the full survey in Appendix~\ref{app:full_operator_survey}, the dimension-nine operators appear to form several characteristic spread patterns within the adopted benchmark setup. This observation should not be regarded as a basis-independent classification of dimension-nine LNV interactions, but it is useful for choosing representative examples. Table~\ref{tab:app_dim9_patterns} summarizes these patterns. The main text keeps $3L^{(9)}$ as a moderate-spread example and includes $6^{(9)}$ as a representative large-spread example.

\begin{table*}[tbp]
\centering
\caption{Characteristic spread patterns of dimension-nine operators in the adopted future benchmark setup. The grouping is only a diagnostic feature of the present calculation and should not be interpreted as a universal classification. The ratio spread refers to the $^{100}$Mo-to-$^{136}$Xe post-discovery comparison.}
\label{tab:app_dim9_patterns}
\begin{tabular}{llcc}
\hline
Pattern & Operators & Benchmark-set spread & Ratio spread $^{100}{\rm Mo}/^{136}{\rm Xe}$ \\
\hline
moderate & $2,3,4,5$-type dimension-nine operators & $\simeq 4.24$ & $\simeq 4.95$ \\
large-spread & $6,7,8,9$-type dimension-nine operators & $\simeq 29.1$ & $\simeq 203$ \\
extreme & $1L,1R$-type dimension-nine operators & $\simeq 108$ & $\simeq 475$ \\
\hline
\end{tabular}
\end{table*}

\section{Full operator survey}
\label{app:full_operator_survey}
\begin{figure*}[t]
  \centering
  \includegraphics[width=0.95\textwidth,height=0.78\textheight,keepaspectratio]{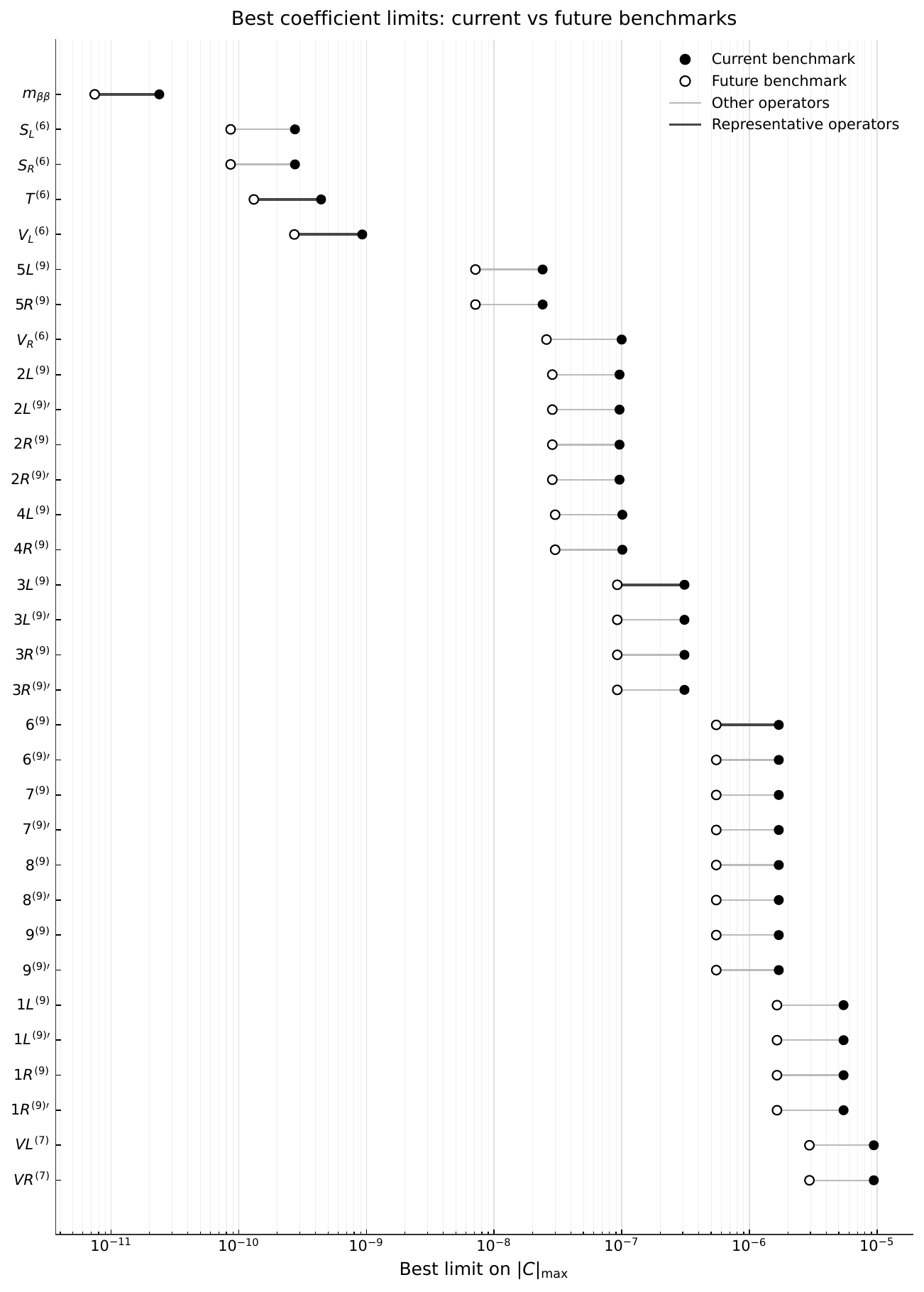}
  \caption{
  Full operator-by-operator counterpart of the left panel of Fig.~\ref{fig:representative_summary}. The figure shows the best coefficient limits for all operator classes in the current and future sets. Filled circles denote the current set and open circles the future set. The highlighted points correspond to the selected operator classes used in the main text.
  }
  \label{fig:app_best_limits}
\end{figure*}

\begin{figure*}[t]
  \centering
  \includegraphics[width=0.95\textwidth,height=0.78\textheight,keepaspectratio]{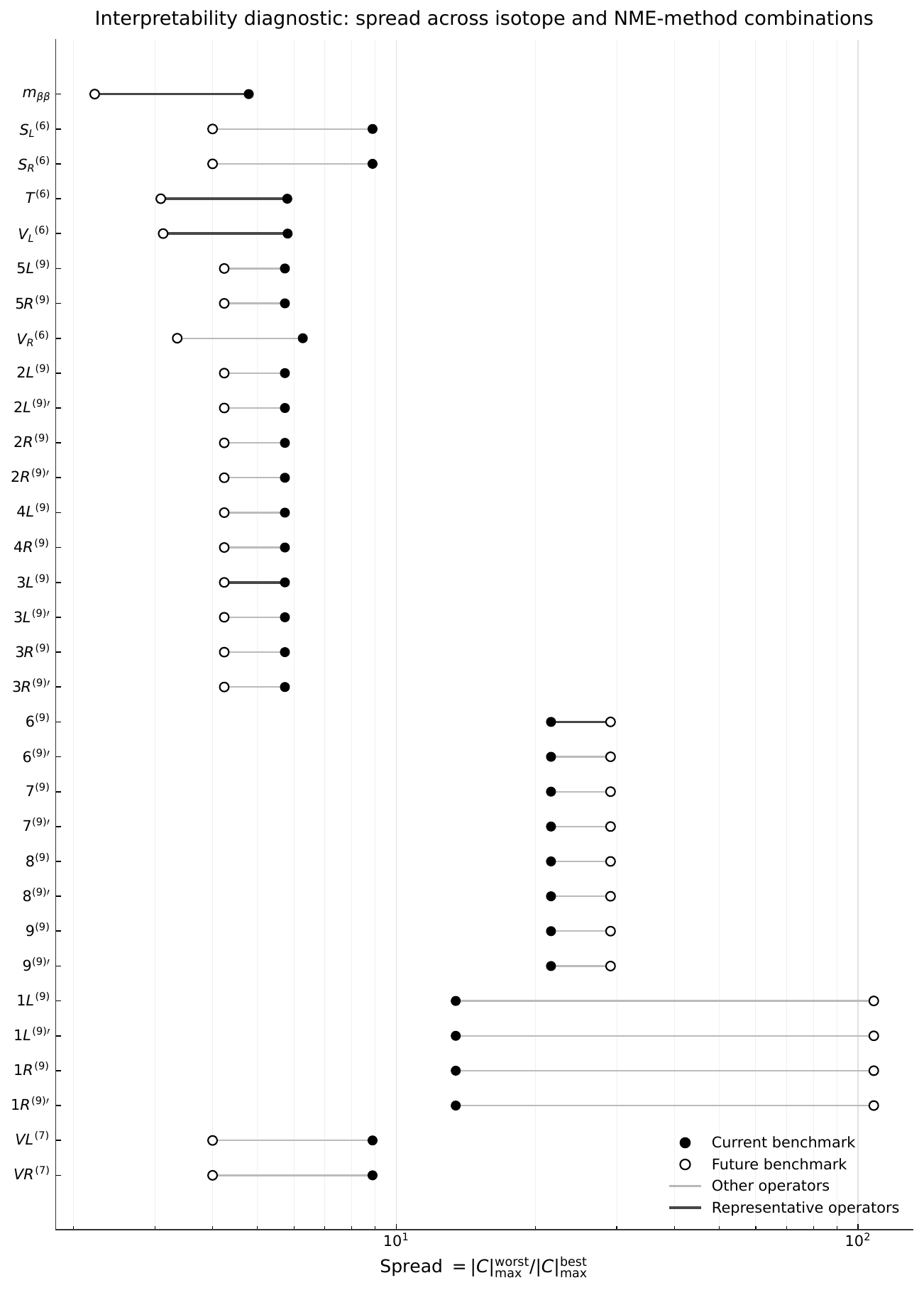}
  \caption{
  Full operator-by-operator counterpart of the right panel of Fig.~\ref{fig:representative_summary}. The figure shows the spread of coefficient limits across isotope and NME-method combinations. The spread is defined as $\mathrm{spread}=|C|_{\max}^{\rm worst}/|C|_{\max}^{\rm best}$. The highlighted points correspond to the selected operator classes used in the main text.
  }
  \label{fig:app_spread}
\end{figure*}

In the main text, Fig.~\ref{fig:representative_summary} focuses on the compact operator set used for the post-discovery analysis. For completeness, Fig.~\ref{fig:app_best_limits} shows the corresponding full operator-by-operator survey of best coefficient limits, and Fig.~\ref{fig:app_spread} shows the full spread distribution across the adopted isotope and NME-method combinations. The decomposition of this benchmark-set spread into NME-only and isotope-only components is given in Appendix~\ref{app:spread_decomposition}.

\clearpage
\bibliography{ref}

\end{document}